\documentclass[aps,prl,reprint,amsmath,amssymb]{revtex4-1}

\newcommand*{\MAINTEXT}{}

\usepackage{graphicx}
\usepackage{dcolumn}
\usepackage{bm}
\usepackage{amsmath}
\usepackage{siunitx}
\usepackage{color}
\usepackage{graphicx}

\begin{document}

\ifdefined\MAINTEXT
\else
	\clearpage
	\setcounter{figure}{0}
	\setcounter{page}{1}
	\renewcommand{\thefigure}{S\arabic{figure}}
\fi

\title{
Low-cost orbital-based linear-scaling \emph{ab initio} molecular dynamics for weakly-interacting systems
}

\author{Hayden Scheiber}
\email{scheiber@chem.ubc.ca}
\affiliation{Department of Chemistry, McGill University, 801 Sherbrooke St. West, Montreal, QC H3A 0B8, Canada}
\author{Yifei Shi}
\affiliation{Department of Chemistry, McGill University, 801 Sherbrooke St. West, Montreal, QC H3A 0B8, Canada}
\author{Rustam Z. Khaliullin}
\email{rustam.khaliullin@mcgill.ca}
\affiliation{Department of Chemistry, McGill University, 801 Sherbrooke St. West, Montreal, QC H3A 0B8, Canada}

\date{\today}

\ifdefined\MAINTEXT

\begin{abstract}
Within the framework of linear-scaling Kohn-Sham density functional theory, a robust method for maintaining compact localized orbitals close to the ground state is coupled with nuclear dynamics. This allows to obviate the commonly employed optimization of the one-electron density matrix and thus create an efficient orbital-only molecular dynamics method for weakly-interacting systems. 
An application to liquid water demonstrates that the low computational overhead of the method makes it well-suited for routine simulations whereas its linear-scaling complexity allows to extend first-principle dynamical studies of molecular systems to previously inaccessible length scales. 
\end{abstract}

\fi

\maketitle

\ifdefined\MAINTEXT


Since the unification of molecular dynamics and density functional theory (DFT)~\cite{a:thecpmd},
\emph{ab initio} molecular dynamics (AIMD) has become an important tool to study processes in molecules and materials.
Unfortunately, the computational cost of the conventional Kohn-Sham (KS) DFT grows cubically with the number of atoms, which severely limits the \emph{length scales} accessible by AIMD. 
To address this issue, substantial efforts have been directed to the development of linear-scaling (LS) DFT.

In all LS DFT methods, the \emph{delocalized} eigenstates of the effective KS Hamiltonian must be replaced with an alternative set of \emph{local} electronic descriptors. 
Most LS methods~\cite{a:ls-rev-1999, a:ls-rev-2012, Kussmann2013, Aarons2016} explore the natural locality of the one-electron density matrix (DM). 
However, the DM DFT becomes advantageous only for impractically large systems when accurate multifunction basis sets are used~\cite{a:ls-dm-sign, Arita2014, a:ls-rev-2012, a:almo-ls}.
This issue is rectified in optimal-basis DM methods~\cite{Skylaris2005, Nakata2015, Mohr2015} that contract large basis sets into a small number of new localized functions and then optimize the DM in the contracted basis. 
Despite becoming the most popular approach to LS DFT, the efficiency of these methods is hampered by the costly optimization of both the contracted orbitals and the DM~\cite{a:ls-onetep-2003}. 
From this point of view, a direct variation of molecular orbitals that are strictly localized within predefined regions is preferable because LS can be achieved with significantly fewer variables. 
Advantages of the orbitals-only LS DFT are especially pronounced in accurate calculations that require many basis functions per atom.
Unfortunately, the development of promising orbital-based LS methods has been all but abandoned~\cite{a:weitao-yang-2013,a:ls-tsuchida-aomm} because of the inherently difficult optimization of localized orbitals~\cite{a:ls-rev-1999,a:ls-mauri-galli-car-1993,a:ls-ordejon-1995,a:ls-fattebert-2004,a:weitao-yang-2013,a:ls-tsuchida-aomm}. 


Thus, despite impressive progress of the LS description of the electronic and atomic structure of large static systems~\cite{Bowler2010,a:ls-dm-sign}, the high computational overhead of existing LS methods restrict their use in dynamical simulations to very short \emph{time scales}, systems of low dimensions, and low-quality minimal basis sets~\cite{a:ls-dm-sign, Otsuka2016, Hine2011, Bowler2010}. 
On typical length and time scales required in practical and accurate AIMD simulations, LS DFT still cannot compete with the straightforward low-cost cubically-scaling KS DFT.

In this work, we present an AIMD method that overcomes difficulties of orbital-only local DFT to achieve LS with extremely low computational overhead. 
To demonstrate advantages of the new method we applied it here to systems of weakly-interacting molecules. 
However, the same approach is readily applicable to systems of strongly-interacting fragments that do not form strong covalent bonds such as ionic materials---salts, liquids, and semiconductors. 
A generalization of the method to all finite-gap systems, including covalently bonded atoms, will be reported later. 

The new AIMD method utilizes a recently developed LS DFT~\cite{a:almo-ls} based on absolutely localized molecular orbitals (\mbox{ALMOs}). 
Unlike delocalized KS orbitals, each \mbox{ALMO} has its own \emph{localization center} and a predefined \emph{localization radius} $R_{c}$ that typically includes nearby atoms or molecules~\cite{a:stoll,a:almo-ls}. 
In the current implementation, a localization center is defined as a set of all Gaussian atomic orbitals of one molecule. 
However, the approach can use other local and nonlocal basis sets~\cite{a:galli_loc, Lin2012}. 
The key feature of ALMO DFT is that its one-electron wavefunctions are constructed in a two-stage self-consistent-field (SCF) procedure~\cite{a:almo-ls} to circumvent the problem of the sluggish variational optimization emphasized above. 
In the first stage, ALMOs are constrained to their localization centers~\cite{a:khal} whereas, in the second stage, ALMOs are relaxed to allow delocalization onto the neighbor molecules within their localization radius $R_{c}$. 
To achieve a robust optimization in the problematic second stage, it is important to keep the delocalization component of the trial wavefunction orthogonal to the fixed orbitals obtained in the first stage. 
For mathematical details, see the ALMO SCF method in Ref.~\citenum{a:almo-ls}.

ALMO constraints imposed by $R_c$ prohibit electron density transfer between distant molecules, but retain all other types of interaction such as long-range electrostatic, exchange, polarization, and---if the exchange-correlation (XC) functional includes them---dispersion interactions~\cite{a:theeda}. 
Since the importance of electron transfer decays exponentially with distance in finite-gap materials~\cite{a:ls-rev-1999}, the \mbox{ALMO} approximation is expected to provide a natural and accurate representation of the electronic structure of molecular systems. 
Because of the greatly reduced number of electronic descriptors and the robust optimization, the computational complexity of ALMO DFT grows linearly with the number of molecules while its computational overhead remains very low. These features make ALMO DFT a promising method for accurate AIMD simulations of large molecular systems.

The challenge of adopting ALMO DFT for dynamical simulations arises from the slightly nonvariational character of the localized orbitals. While ALMOs are variationally optimized in both SCF stages, the occupied subspace defined in the first stage must remain fixed during the second stage to ensure convergence. In addition, electron transfer effects can suddenly become inactive in the course of a dynamical simulation when a neighboring molecule crosses the localization threshold $R_{c}$. Futhermore, the variational optimization in any AIMD method is never complete in practice and interrupted once the maximum norm of the gradient of the energy with respect to the electronic descriptors drops below small but nevertheless finite convergence threshold $\epsilon_{\text{SCF}}$. These errors do not affect the accuracy of static ALMO DFT calculations, geometry optimization, and Monte-Carlo simulations. Unfortunately they tend to accumulate in molecular dynamics trajectories leading to non-physical sampling and eventual failure. 
Traditional strategies to cope with these problems are computationally expensive and include computing the nonvariational contribution to the forces via a variational coupled-perturbed procedure~\cite{Kussmann2013,Benoit2001}, increasing $R_c$, and decreasing $\epsilon_{\text{SCF}}$. 

In this work, we propose another approach that obviates the need in a coupled-perturbed solver, relaxes tight constraints on $R_{c}$ and $\epsilon_{\text{SCF}}$, and thus enables us to maintain stable dynamics and to keep the algorithmic complexity and cost of simulations low. 
In our approach, the forces on atoms are calculated \emph{approximately} after the two-stage ALMO SCF using a straightforward procedure that computes only the Hellmann-Feynman and Pulay components and neglects the computationally intense nonvariational component of the forces. 
The difference between these approximate ALMO forces and the \emph{reference} forces that could be obtained from perfectly converged fully-delocalized KS orbitals is $\delta f_{i\alpha}(t)$:
\begin{align}
\label{eq:deltaf}
f^{\text{KS}}_{i\alpha}(t) = f^{\text{ALMO}}_{i\alpha}(t) + \delta f_{i\alpha} (t),
\end{align}
where $\alpha$ is a Cartesian component of the force acting on atom $i$ at time $t$. $\delta f_{i\alpha} (t)$ comprises all neglected terms that originate from a finite localization radius $R_c$ and incomplete SCF optimization. 
$\delta f_{i\alpha} (t)$ can be reduced to zero \emph{systematically} by increasing $R_c$ and decreasing $\epsilon_{\text{SCF}}$.

Our approach to compensate for the missing $\delta f_{i\alpha}(t)$ term is inspired by the methodology introduced into AIMD by Krajewski \emph{et al.}~\cite{Krajewski}, formalized by K\"uhne \emph{et al.}~\cite{a:2ndcpmd} and rationalized by Dai \emph{et al.}~\cite{a:langevin-why} before becoming informally known as the second generation Car-Parrinello molecular dynamics~\cite{Kuhne2013}. 
Adopting the principle of Refs.~\onlinecite{Krajewski} and \onlinecite{a:2ndcpmd}, ALMO AIMD is chosen to be governed by the Langevin equation of motion that can be written in terms of the unknown reference forces
\begin{align}
\label{eq:langevin}
m_i \ddot{r}_{i\alpha} = f^{\text{KS}}_{i\alpha}(t) - \gamma m_i \dot{r}_{i\alpha} + R^{\gamma}_{i\alpha} (t),
\end{align}
where $m_i$ is the mass of atom $i$, $r_{i\alpha}$ is its position along dimension $\alpha$, $\gamma$ is the Langevin scaling factor, and $R^{\gamma}_{i\alpha} (t)$ is the stochastic force represented by a zero-mean white Gaussian noise 
\begin{align}
\label{eq:stochastic}
\langle R^{\gamma}_{i\alpha} (t) \rangle &= 0, \\
\langle R^{\gamma}_{i\alpha} (t)  R^{\gamma}_{j\beta} (t') \rangle &= 2 k_B T \gamma m_i \delta_{ij} \delta_{\alpha\beta} \delta(t-t').
\end{align}
The last relation means that, for any value of $\gamma$, the damping and stochastic terms are in perfect balance and trajectories generated with Eq.~(\ref{eq:langevin}) will sample the canonical ensemble at a specified temperature $T$~\cite{a:Kubo-1986}. 
In the limit $\gamma \rightarrow 0$, the Newton equation is recovered and the microcanonical ensemble is sampled. 

The main assumption of ALMO AIMD is that the error in the ALMO forces is well approximated by Gaussian noise $R^{\Delta}_{i\alpha} (t)$:
\begin{align}
\label{eq:assumption}
\delta f_{i\alpha}(t) = R^{\Delta}_{i\alpha} (t)
\end{align}
that obeys
\begin{align}
\label{eq:stochastic2}
\langle R^{\Delta}_{i\alpha} (t) \rangle &= 0, \\
\label{eq:stochastic3}
\langle R^{\Delta}_{i\alpha} (t)  R^{\Delta}_{j\beta} (t') \rangle &= 2 k_B T \Delta m_i \delta_{ij} \delta_{\alpha\beta} \delta(t-t') .
\end{align}
This assumption, shown to be well justified, allows us to rewrite the Langevin equation using the ALMO forces
\begin{align}
\label{eq:langevin2}
m_i \ddot{r}_{i\alpha} = f^{\text{ALMO}}_{i\alpha}(t) - \gamma m_i \dot{r}_{i\alpha} + R^{\gamma + \Delta}_{i\alpha} (t),
\end{align}
where the two stochastic terms are combined into one $R^{\gamma + \Delta}_{i\alpha} = R^{\gamma}_{i\alpha} + R^{\Delta}_{i\alpha}$. 
The only missing piece in the modified Langevin equation is the value of $\Delta$, which describes the strength of the newly introduced stochastic term. 
This term compensates for imperfections in ALMO forces and must be adjusted to re-balance the damping and stochastic components in ALMO AIMD.



In principle, $\Delta$ can be calculated using the integral of Eq.~(\ref{eq:stochastic3}) averaged over atoms with different $m_i$
\begin{align}
\label{eq:delta}
\Delta &= (2 k_B T m_i )^{-1} \int_{-\infty}^{\infty} \frac{1}{3} \langle \delta \vec{f}_i (0) \cdot \delta\vec{f}_i(\tau) \rangle d\tau
\end{align}
if one can afford computing the reference forces $f^{\text{KS}}_{i\alpha}(t)$ (i.e. $R_c \rightarrow \infty$ and $\epsilon_{\text{SCF}} \rightarrow 0$) for a short representative AIMD trajectory. 
In practice, we found (see results below) that this approach is not particularly accurate because the $\delta_{ij}\delta_{\alpha\beta}$ assumption in Eq.~(\ref{eq:stochastic3}) does not strictly hold. 
Nevertheless, the ACF integral can provide a reasonable starting value of $\Delta$. This value can be further fine-tuned in a series of short trial-and-error ALMO AIMD runs until the average kinetic energy corresponds to the requested temperature $\langle \frac{1}{2} m_i \dot{\bm{r}}^{2}_{i} \rangle = \frac{3}{2} k_{B} T$.

The inherently stochastic approach presented here does not aim to produce fully time-reversible dynamics for atomic nuclei. 
Nevertheless, it is capable to reproduce correct dynamical properties of a system as long as $\gamma$ is set to a small value and  partially optimized ALMOs remain close to the ground state resulting in $\Delta \ll \gamma$. 


\begin{figure}
\includegraphics[trim={1.3cm 0.1cm 3.3cm 1.3cm},clip,width=8.6cm]{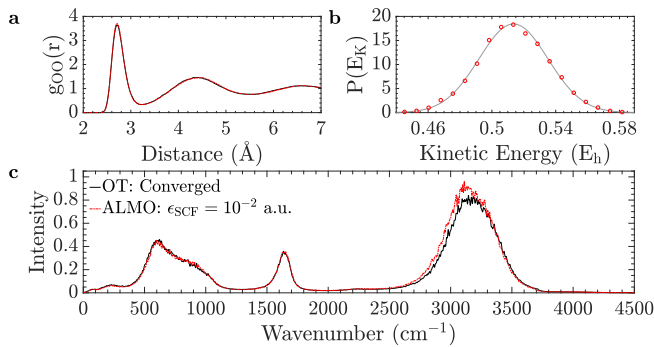}
\caption{\label{fig:dynproperties} 
Calculated properties of water using ALMO AIMD with $\epsilon_{\text{SCF}} = 10^{-2}$~a.u. and $R_{c} = 1.6$ vdWR (red line) and fully converged OT reference (black line).
(a) RDF, 
(b) kinetic energy distribution (the gray curve shows the theoretical Maxwell-Boltzmann distribution), 
(c) IR spectrum.
}
\end{figure}

ALMO AIMD was implemented in CP2K, an open source materials modeling package~\cite{www:cp2k}. 
Accuracy and efficiency of ALMO AIMD was tested using liquid water as an example. 
This system is challenging because intermolecular electron delocalization is a critical component of hydrogen bonding and must be described correctly to reproduce static and dynamical properties of liquid water. 
A periodic cell containing 125 molecules was simulated for $\SI{30}{\ps}$ at $T=\SI{298}{\K}$ and a constant density of $\SI{1.01}{\g\cdot\cm^{-3}}$. 
Ricci-Ciccotti algorithm~\cite{Ricci2003} was used to integrate the Langevin equation. We found that $\gamma = \SI{e-3}{\per\fs}$ is large enough to thermostat the system efficiently and small enough not to significantly affect dynamical properties of liquid water.
In the dual Gaussian and plane-wave scheme implemented in CP2K~\cite{a:quickstep}, the TZV2P basis set was used to represent molecular orbitals, and a plane-wave cutoff of $\SI{320}{Ry}$ used to represent electron density. 
The XC energy was approximated using the dispersion-corrected PBE functional~\cite{a:pbe,Grimme2010}. 
Separable norm-conserving pseudopotentials were used~\cite{a:hgh} and the Brillouin zone was sampled at the $\Gamma$-point. 
The predictor of the Kolafa scheme~\cite{Kolafa2003} was adopted to localized orbitals~\cite{a:2ndcpmd} to generate a highly accurate initial ALMOs in both SCF stages, which can be brought close to the ground state with just a few SCF steps of the robust two-stage optimization procedure. 

The \emph{reference} forces were calculated with fully delocalized electrons using the tightly converged, $\epsilon_{\text{SCF}}=10^{-6}$~a.u., orbital transformation (OT) method~\cite{a:ot}. 
In ALMO AIMD, the element-specific cutoff radius for electron delocalization $R_c$ was set to 1.6 in units of the elements' van der Waals radii (vdWR). This localization radius includes approximately two coordination shells of an average water molecule and was shown to reproduce the reference radial distribution function (RDF) perfectly in Monte-Carlo simulations~\cite{a:almo-ls}. 
To check the ability of the $R^{\Delta}(t)$ term to compensate for imperfections in ALMO forces, we varied $\epsilon_{\text{SCF}}$ between tight $10^{-6}$~a.u. and loose $10^{-2}$~a.u. 

\begin{figure}
\includegraphics[trim={0.5cm 0cm 0.7cm 0.1cm},clip,width=8.6cm]{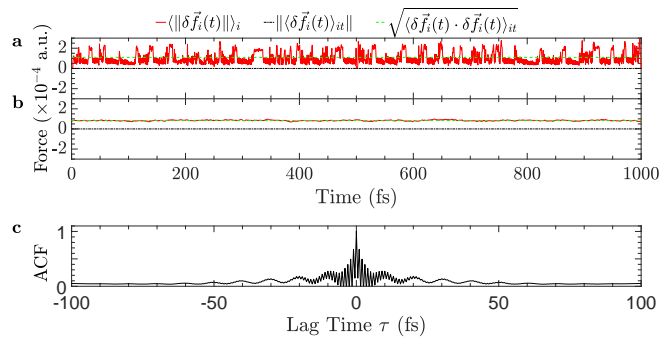}
\caption{\label{fig:randomforce} 
The red line is the Euclidean norm of the instantaneous error $\langle \| \delta \vec{f_{i}}(t) \| \rangle_{i}$, the black line is the magnitude of the time average of the instantaneous error vector, and the green line is the time average of the red line. 
(a) $R_{c} = 1.6$ vdWR and $\epsilon_{\text{SCF}} = 10^{-2}$~a.u., 
(b) $R_{c} = 1.6$ vdWR and fully converged ALMO SCF, 
(c) Normalized ACF $\frac{1}{3}\langle \delta \vec{f}_i (t) \cdot \delta\vec{f}_i(t+\tau) \rangle_{it}$ of the instantaneous error in panel (a).
}
\end{figure}

Even with $\epsilon_{\text{SCF}} = 10^{-2}$, the simulation is stable with the correct average temperature and perfect Maxwell-Boltzmann distribution (Figure~\ref{fig:dynproperties}b). $\Delta$ was initially estimated at $\SI{2e-5}{\per\fs}$ using Eq.~(\ref{eq:delta}) and then refined heuristically to $\SI{6e-5}{\per\fs}$. 
We found that it is easier to optimize $\Delta$ when $\gamma$ is set to zero because of reduced noise in the trial runs. 
%
Analysis of $\delta \vec{f_{i}}(t)$ shows that the error indeed resembles Gaussian white noise. The mean of the error is zero (black line in Figure~\ref{fig:randomforce}a). Its ACF decays rapidly (Figure~\ref{fig:randomforce}c) so that the errors can be considered uncorrelated on time scale of $\SI{50}{\fs}$. Thus the main assumption behind our approach to ALMO AIMD is justified for liquid water. We established that the main source of error in forces for this system is the loose convergence criterion and not the finite $R_c$: fully converged ALMO SCF calculations remove the oscillating component of $\delta f$ (Figure~\ref{fig:randomforce}b). We also verified that the ALMO forces converge to the reference forces in the limit $R_{c} \rightarrow \infty$ 
(Figure~S1 in the Supplemental Material).



To test the accuracy of ALMO AIMD we used the trajectory analyzer TRAVIS~\cite{a:travis-main} to compute the infrared (IR) spectrum, RDF, and diffusion coefficient of liquid water from both the ALMO trajectory ($\epsilon_{\text{SCF}} = 10^{-2}$~a.u. and $R_{c} = 1.6$ vdWR) and from the reference trajectory. 
The diffusion coefficients $D_{\text{OT}}=\SI{1.7\pm0.1E-10}{\m^{2}\cdot\s^{-1}}$ and $D_{\text{ALMO}}=\SI{1.8\pm0.4E-10}{\m^{2}\cdot\s^{-1}}$ and RDFs (Figure~\ref{fig:dynproperties}a) are in good agreement. The quality of the ALMO IR spectrum (Figure~\ref{fig:dynproperties}c) is good despite minor errors in the intensity of the OH stretching mode, which is sensitive to the precise positions of the centers of localized orbitals. These stringent tests show that despite noticeable errors in the ALMO forces (Figure \ref{fig:randomforce}a), the compensating $R^{\Delta}(t)$ term in the modified Langevin equation makes it possible to recover atomic dynamics properly. We would like to note that ALMO AIMD could not be stabilized with $\Delta=0$. Neither were we able to find any values of $\Delta$ that stabilize trajectories generated using perturbative versions of ALMO DFT~\cite{a:almo-ls}.



To demonstrate the computational efficiency of ALMO AIMD, we compared the average wall-time per MD step for a variety of methods in Figure \ref{fig:strongscaling_log}.
It is important to emphasize that the comparison is performed for a three-dimensional condensed phase system described with an accurate triple-$\zeta$ basis set with polarized functions---a particularly challenging case for DM-based LS methods.
ALMO AIMD shows clear LS behavior for all values of $\epsilon_{\text{SCF}}$, even for medium-size systems. While the second generation Car-Parrinello method decreases the computational overhead of the cubically-scaling AIMD for small systems~\cite{a:2ndcpmd}, ALMO AIMD exploits the modified Langevin concept to substantially reduce the simulation cost for systems of all sizes.
The crossover point between ALMO AIMD and cubically scaling methods lies in the region of 256 molecules---length scales routinely accessible with AIMD today. 


\begin{figure}
\includegraphics[trim={2.5cm 0.5cm 3.4cm 0.9cm},clip,width=8.6cm]{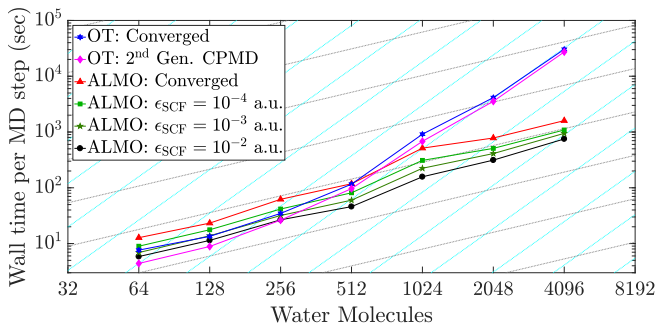}
\caption{\label{fig:strongscaling_log} Timing benchmarks for PBE/TZV2P simulations of liquid water on 256 compute cores. 
For ALMO methods, $R_{c} = 1.6$ vdWR. 
Cyan lines represent perfect cubic scaling, whereas gray lines represent perfect linear scaling. 
}
\end{figure}


Weak scaling benchmarks for very large systems show (Figure~\ref{fig:weakscaling}) that localized orbitals are naturally suited for parallel execution: LS is retained for a wide range of systems and compute cores. 
We were able to successfully simulate systems as large as $\sim10^{5}$ atoms 
within reasonable wall-clock time using only moderate number of compute cores---an impressive feat for AIMD considering that accurate molecular orbitals and the idempotent DM are computed on each step.
The horizontal line in Figure~\ref{fig:weakscaling} is shown as a rough guide to time and length scales accessible in a fixed wall-clock time given various computational resources. It indicates that ALMO AIMD can extend the range of routine simulations to $\sim10^4$ atoms on modern HPC platforms. 



\begin{figure}
\includegraphics[trim={1.6cm 0.8cm 4.7cm 0cm},clip,width=8.6cm]{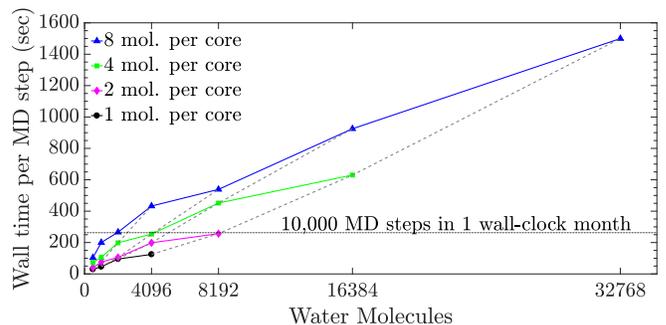}
\caption{\label{fig:weakscaling} Weak scalability benchmarks for PBE/TZV2P ALMO AIMD with  $R_{c} = 1.6$ vdWR and $\epsilon_{\text{SCF}} = 10^{-2}$~a.u. 
Dashed gray lines connect systems simulated on the same number of cores to confirm LS behavior. 
}
\end{figure}


To summarize, we demonstrated---for the first time---that compact localized orbitals can be utilized to perform accurate and efficient LS AIMD without concomitant optimization of the DM. 
High efficiency of the presented method is achieved without sacrificing accuracy with a combination of two techniques: (1) on-the-fly calculation of approximate forces without lengthy self-consistent optimization of localized orbitals and (2) integration of a modified Langevin equation of motion that is fine-tuned to retain stable dynamics even with imperfect forces. 
By obviating the optimization of the DM, the method remains remarkably efficient even with large localized basis sets. 
Using liquid water as an example, we showed that the new approach enables simulations of molecular systems on previously inaccessible length scales. 
The developed method will have a significant impact on modeling of complex molecular systems (e.g. interfaces or nuclei) making completely new phenomena accessible to AIMD. 
Generalization of the methodology to systems of strongly interacting atoms (e.g. covalent crystals) is underway.

\textbf{Acknowledgments.} The authors are grateful to Thomas K\"uhne for insightful discussions. The research was funded by the Natural Sciences and Engineering Research Council of Canada through the Discovery Grant (RGPIN-2016-05059). The authors are grateful to Compute Canada and McGill HPC Centre for computer time.

%


	\clearpage
	\setcounter{figure}{0}

	\renewcommand{\thefigure}{S\arabic{figure}}
\subsection{Supplementary material: Dependence of the forces on the electron localization radius}

The ALMO forces converge to the reference forces as the electron localization radius increases (Figure~\ref{fig:forcecomp}).

\begin{figure}[h!]
\includegraphics[trim={0cm 0cm 0.1cm 0.1cm},clip,width=8.6cm]{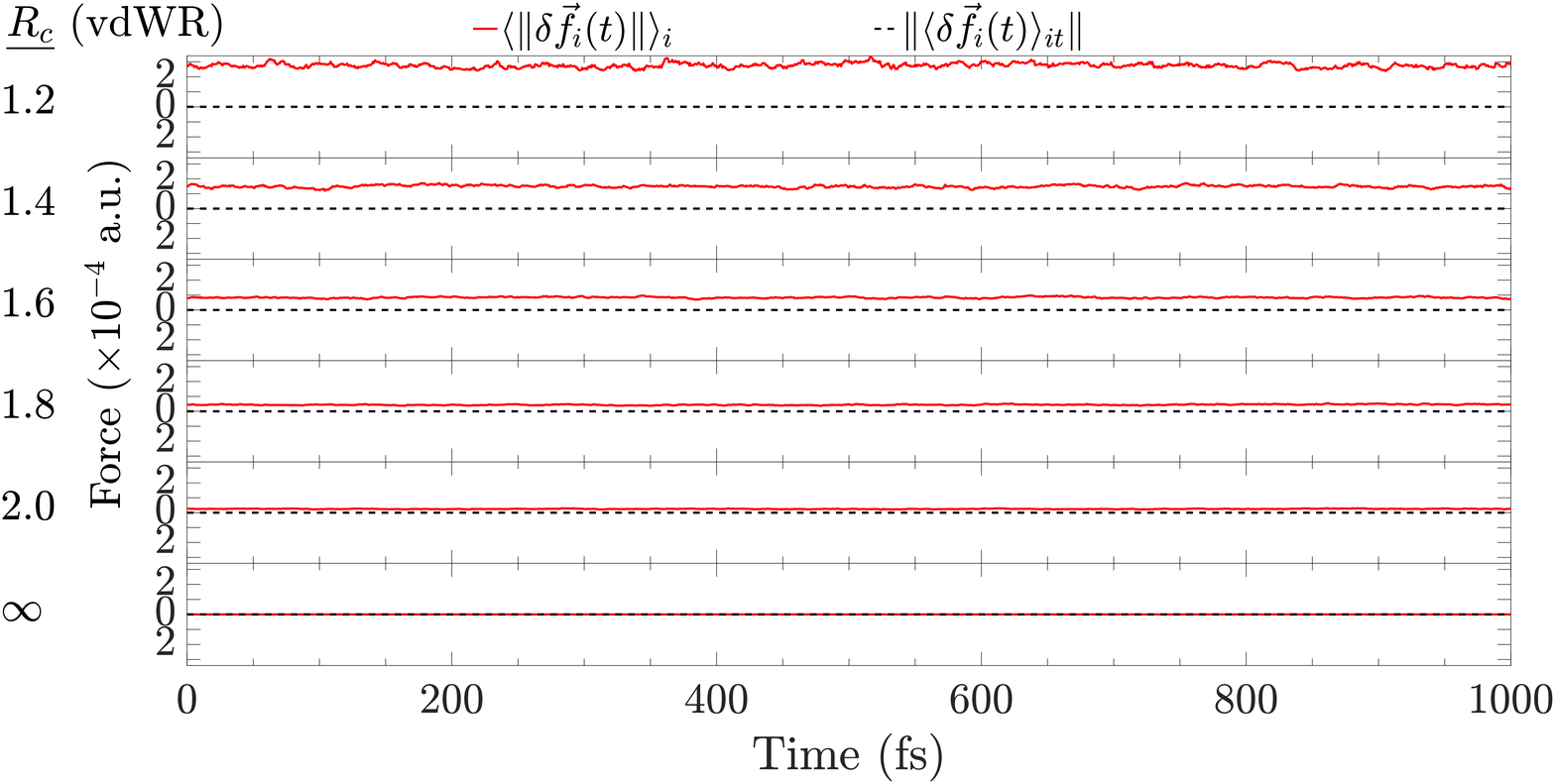}
\caption{\label{fig:forcecomp} Dependence of the Euclidean norm of the instantaneous error $\langle \| \delta \vec{f_{i}}(t) \| \rangle_{i}$ (red line) on the ALMO localization radius $R_c$ expressed in units of atomic van der Waals radii. Black line is the magnitude of the time average of the instantaneous error vector. 
Forces are computed with the PBE XC functional, TZV2P basis set, and $\epsilon_{\text{SCF}} = 10^{-7}$~a.u. for the configurations from a 30-ps trajectory generated with delocalized AIMD at T=298~K.}
\end{figure}


\else
\fi 

\end{document}